\documentstyle[12pt,aaspp4]{article}

\def\Ms {$M_{\ast}$}
\def\Rs {$R_{\ast}$}
\def\degr{$^{\circ}$}
\def\hd{HD~209458}
\def\Teff{$T_{\rm eff}$}
\def\logg{$\log g$}
\def\Mv{$M_V$}
\newcommand\th{\thinspace}
\newcommand\kms{\ifmmode{\rm km\th s^{-1}}\else km\th s$^{-1}$\fi} 
\newcommand\ms{\ifmmode{\rm m\th s^{-1}}\else m\th s$^{-1}$\fi}
\newcommand\msun{\ifmmode{M_{\odot}}\else $M_{\odot}$\fi}
\newcommand\MJ{\ifmmode{M_{Jup}}\else $M_{Jup}$\fi}
\newcommand\RJ{\ifmmode{R_{Jup}}\else $R_{Jup}$\fi}
\newcommand\Mo{\ifmmode{M_{\odot}}\else $M_{\odot}$\fi}
\newcommand\rsun{\ifmmode{R_{\odot}}\else $R_{\odot}$\fi}
\newcommand\Ro{\ifmmode{R_{\odot}}\else $R_{\odot}$\fi}

\begin{document}

\title{The Spectroscopic Orbit of the Planetary Companion 
Transiting HD~209458\altaffilmark{1}}

\author{Tsevi Mazeh,\altaffilmark{2}
Dominique Naef,\altaffilmark{3}
Guillermo Torres,\altaffilmark{4}
David W. Latham,\altaffilmark{4}
Michel Mayor,\altaffilmark{3}
Jean-Luc Beuzit,\altaffilmark{5}
Timothy M. Brown,\altaffilmark{6}
Lars Buchhave,\altaffilmark{7}
Michel Burnet,\altaffilmark{3}
Bruce~W.~Carney,\altaffilmark{8}
David Charbonneau,\altaffilmark{4,6}
Gordon A. Drukier,\altaffilmark{2}
John B. Laird,\altaffilmark{9}
Francesco~Pepe,\altaffilmark{3}
Christian Perrier,\altaffilmark{5}
Didier Queloz,\altaffilmark{3}
Nuno C. Santos,\altaffilmark{3}
Jean-Pierre Sivan,\altaffilmark{10}
St\'ephane~Udry,\altaffilmark{3}
Shay Zucker\altaffilmark{2}}

\altaffiltext{1}{Some of the data presented herein were obtained at the
W.\ M.\ Keck Observatory, which is operated as a scientific partnership
among the California Institute of Technology, the University of
California and the National Aeronautics and Space Administration.  The
Observatory was made possible by the generous financial support of the
W.M. Keck Foundation. The other data were obtained at Observatoire de
Haute-Provence (France) and with the 1.2-m Euler Swiss telescope at La
Silla Observatory, ESO Chile.}

\altaffiltext{2}{School of Physics and Astronomy, Raymond and Beverly 
Sackler Faculty of Exact Sciences, Tel Aviv University, Tel Aviv 69978,
Israel}

\altaffiltext{3}{Observatoire de Gen\`eve, 51 ch. des Maillettes, CH
  1290 Sauverny, Switzerland}

\altaffiltext{4}{Harvard-Smithsonian Center for Astrophysics, 60
  Garden Street, Cambridge, MA 02138}

\altaffiltext{5}{Laboratoire d'Astrophysique, Observatoire de
  Grenoble, BP 53, F-38041 Grenoble, France}

\altaffiltext{6}{High Altitude Observatory, National Center for
  Atmospheric Research, P.O. Box 3000, Boulder, CO 80307-3000.
  NCAR is sponsored by the National Science Foundation.}

\altaffiltext{7}{Astronomical Observatory 
Niels Bohr Institute for Astronomy, Physics \& Geophysics
Copenhagen University,
Denmark}

\altaffiltext{8}{Department of Physics and Astronomy, University of
  North Carolina at Chapel Hill, Chapel Hill, NC 27599-3255}

\altaffiltext{9}{Department of Physics \& Astronomy, Bowling Green
  State University, Bowling Green, Ohio 43403}

\altaffiltext{10}{Observatoire de Haute-Provence, F-04870, St-Michel
  l'Observatoire, France}

\begin{abstract}
  
  We report a spectroscopic orbit with period $P = 3.52433\pm 0.00027$
  days for the planetary companion that transits the solar-type star
  HD~209458. For the metallicity, mass, and radius of the star we derive
  [Fe/H]$ = 0.00 \pm 0.02$, $M_* = 1.1 \pm 0.1\, \msun$, and $R_* = 1.3
  \pm 0.1 \,\rsun$. This is based on a new analysis of the iron lines in our HIRES
  template spectrum, and also on the absolute magnitude and color of the
  star, and uses isochrones from four different sets of stellar evolution
  models. Using these values for the stellar parameters we reanalyze the
  transit data and derive an orbital inclination of $i = 85\fdg2 \pm
  1\fdg4$. For the planet we derive a mass of $M_{p} = 0.69 \pm 0.05\, 
  M_{\rm Jup}$, a radius of $R_{p} = 1.54 \pm 0.18\, R_{\rm Jup}$, and a
  density of $\rho = 0.23 \pm 0.08 \, {\rm g \, cm^{-3}}$.
 
\subjectheadings{binaries: eclipsing --- planetary systems --- 
stars: individual (HD 209458) ---
techniques: radial velocities}

\end{abstract}

\section{INTRODUCTION}

We report in this Letter on our spectroscopic observations of
HD~209458, observations that led to the discovery of a transiting
planet with an orbital period of 3.5 days.
 
We have been observing HD~209458 since August 1997 as one of the targets
in two large independent radial-velocity surveys, both searching for
extrasolar planets around solar-type stars. One program uses HIRES (Vogt
et al.\ 1994) on the Keck I telescope, and the other uses ELODIE (Baranne
et al.\ 1996) on the 1.93-m telescope at Observatoire de Haute Provence
(France). In June 1999, after observations from both efforts showed that
the radial velocity of HD~209458 was variable, additional frequent
observations were obtained with ELODIE, as well as with CORALIE (Queloz et
al.\ 1999) on the new 1.2-m Swiss telescope at La Silla.
 
In August 1999 the identity of HD~209458 and its orbital elements were
provided to D.\ Charbonneau and T.\ Brown so that they could look for
transits with the STARE photometric instrument (Brown \& Kolinski 1999).
Two transits were successfully observed in September 1999 (Charbonneau
et al.\ 2000; hereafter C00).  An independent discovery in November
1999 of the planetary orbit, as well as the detection of a transit ingress, are
reported by Henry et al.\ (2000).
 
Transit observations together with an orbital solution allow us to
determine directly the mass, radius, and density of the planet, provided
we have good estimates for the mass, radius, and limb darkening of the
star (e.g.,\ C00). We describe in this Letter our
efforts to derive better estimates for these parameters.

\section{OBSERVATIONS}

One of the two radial-velocity projects, the results of which we report
here, is the G Dwarf Planet Search (Latham 2000). The sample for
this project is composed of more than 1000 targets whose effective
temperatures, luminosities, chemical compositions, and
Galactic-population memberships have been determined using precise
Str\"omgren photometry (Olsen 1993; private communications).  In
addition, the radial velocities of these stars were known to be constant
at a precision of 300--600 \ms, based on more than ten years of
observations with the CORAVELs (Mayor 1985) and the CfA Digital Speedometers
(Latham 1985, 1992).

The observations for this project were performed with HIRES and its
iodine gas-absorption cell (Marcy \& Butler 1992) on the Keck I
telescope.  The G Dwarf Planet Search observing strategy is designed to
carry out an initial reconnaissance of the sample stars, with the
immediate goal of identifying the stars whose radial velocity is
modulated with an amplitude of about 50 \ms\ or larger.  To increase the
number of target visits we concede velocity precision, and therefore have
exposed no longer than needed to achieve a precision of 10 \ms. Radial
velocities are derived from the spectra with TODCOR (Zucker and Mazeh
1994) --- a two dimensional correlation algorithm.  Although we are still
in the development stage of our software (Zucker, Drukier \& Mazeh, in
preparation), preliminary results from the global analysis of the 675
stars with two or more iodine observations show that we are already
close to achieving our goal of 10 \ms\ or better.

The other program whose results are presented here is the ELODIE planet
search survey (Mayor \& Queloz 1995a). After the discovery of the
planetary companion around 51 Peg (Mayor \& Queloz 1995b), the surveyed
sample was extended to about 320 northern hemisphere solar-type stars
brighter than $m_V=7.65$ and with small projected rotational velocities
($v\sin i$ from CORAVEL, Benz \& Mayor 1984). From CORAVEL data, the
stars in this sample were known to have constant radial velocities at a
300 \ms\ precision level.  Radial-velocity measurements are obtained
with the ELODIE echelle fiber-fed spectrograph (Baranne et al. 1996)
mounted on the Cassegrain focus of the 1.93-m telescope of the
Observatoire de Haute-Provence. The reduction technique used for this
sample is known as the ``simultaneous Thorium-Argon technique" described
by Baranne et al (1996).  The precision achieved with this instrument is
of the order of 10 \ms\ over more than 3 years.

After the two independent detections of the short-term variability of
\hd\ by the HIRES and ELODIE teams, we decided to add this object to the
CORALIE planet search sample (Udry et al.  1999a,b) in order to gather
more radial-velocity data and therefore increase the precision of the
orbital elements.  The precision achieved with
CORALIE is of the order of 7--8 \ms\ over 18 months.  To check for other
possible sources of line shifts besides orbital motion we computed the
mean bisector profiles (as described by Queloz et al.\ 2000, in
preparation) for all the ELODIE and CORALIE spectra. No correlation
between the observed velocities and the line profiles was detected,
convincing us that the planetary interpretation was correct even before
transits were detected.

\section{RADIAL-VELOCITY ANALYSIS}

As of November 16, 1999, we had a total of 150 radial-velocity
measurements of HD~209458 available for analysis: 11 from HIRES, 31 from
ELODIE, and 108 from CORALIE. Initially we applied shifts of $-5$
m~s$^{-1}$ to the ELODIE velocities and $-14780$ m~s$^{-1}$ to the HIRES
measurements to bring them to the CORALIE system, the latter offset
being much larger due to the arbitrary zero point of the HIRES
velocities (Zucker, Drukier \& Mazeh, in preparation). To account for
possible errors in these shifts, the orbital solutions described below
included two additional free parameters --- $\Delta_{H-C}$ and
$\Delta_{E-C}$ for the HIRES and ELODIE shifts --- along with the orbital
elements.

In addition, the two transit timings recorded by C00 provide useful
information on the orbital period and $T_c$, the time of inferior
conjunction.  These timings were therefore included in the derivation
of the spectroscopic orbital elements, and we treated them as
independent measurements with their corresponding uncertainties.

In a preliminary solution weights were assigned to each observation
based on the internal errors.  From this fit we computed the RMS
residuals separately for each data set -- $\sigma_H$, $\sigma_E$, and
$\sigma_C$ --- and then scaled the internal errors for each instrument to
match the corresponding RMS residuals on average. We re-solved for the
orbital parameters, and the procedure converged in one iteration with
essentially no change in the elements. 

Tests allowing for a non-circular Keplerian orbit for HD~209458 resulted
in an eccentricity indistinguishable from zero: $e = 0.016 \pm 0.018$.
We therefore assume in the following that the orbit is circular. Our
final orbital solution is represented graphically in Figure 1. The
elements are given in Table 1, where the value of the planetary mass,
$M_p$, depends on the inclination angle and on the adopted stellar mass,
\Ms, to the power of 2/3.  The orbital elements reported by Henry et al.
(2000) are consistent with our results, although their
quoted errors are substantially larger.  Robichon \& Arenou (2000) have
identified three transits in the Hipparcos photometry and have derived a
more precise period, $P = 3.524739 \pm 0.000014$ d, consistent with the value
of Table 1 within 1.5-$\sigma$.

\section{STELLAR PARAMETERS}

In this section we compare the results of two different approaches for
estimating the mass and radius of HD~209458.  In the first approach we
rely on the effective temperature, $T_{\rm eff}$, and the surface
gravity, $\log g$, derived from a fine analysis of the iron lines in the
HIRES template spectrum of \hd.  In the second approach we take
advantage of the accurate distance available from the Hipparcos
astrometric mission (ESA 1997) and rely on the stellar absolute magnitude and
observed color. For both approaches we matched theoretical isochrones
from four different sets of stellar evolution models with the location of
\hd\ in the corresponding parameter plane, estimating the stellar mass,
radius and age.

The stellar models depend sensitively on the metallicity, and therefore
a critical first step for both approaches is to determine the
metallicity. 
An analysis of eight spectra obtained with the CfA Digital Speedometers
(Latham 1992) with the techniques reported by Carney et al.\ (1987) gave
$T_{\rm eff}=5975$ K, $\log g = 4.25$, and ${\rm [m/H]}=+0.11 \pm 0.1$.
Another independent analysis of the cross-correlation dips from CORAVEL
observations (Mayor 1980 as revised by Pont 1997 using primary
calibrators by Edvardsson et al. 1993) gave [Fe/H]$=-0.14 \pm 0.1$.

This large range in metallicity values, $-0.14$ to $+0.11$, implied a
large uncertainty in the mass and radius of HD~209458, so we undertook a
detailed analysis of selected Fe I and Fe II lines measured on our HIRES
template spectrum, which has a resolving power of about 70,000 and
signal-to-noise ratio per resolution element of about 300. We adopted a
line list developed by L.\ de Almeida (private communication) and solar
{\it gf} values based on the National Solar Observatory solar flux atlas
(Kurucz et al.\ 1984).

For this analysis of HD~209458 we used model atmospheres and computer codes
based on the work of R.\ Kurucz. Selected weak Fe I lines were used to
set \Teff\ by adjusting the \Teff\ until the plot of abundance versus
excitation potential was flat.
Stronger Fe I lines were then included and the microturbulent velocity
was adjusted to get a flat dependence of abundance on line strength.
Finally, the surface gravity was adjusted until the abundances from the
Fe II and Fe I lines agreed. This analysis gave $T_{\rm eff}=6000$ K,
microturbulent velocity $\xi = 1.15\, \kms$, $\log g=4.25$, and
[Fe/H]$=0.00$.  The errors in these values are undoubtedly dominated
by systematic effects, and we estimate that they are $\pm 50$ K in
$T_{\rm eff}$, $\pm 0.2$ in $\log g$, and $\pm 0.02$ in [Fe/H]. 

This approach yielded \Teff\ and \logg, as well as the metallicity. In the
other approach we used the absolute magnitude and color of $M_V=4.28 \pm
0.11$ and $B-V=+0.594 \pm 0.015$ (assuming no extinction or reddening),
together with the metallicity derived  from the iron lines.

In Table 2 we compare the values of the stellar mass and radius, \Ms\
and \Rs, and age that we derive for \hd\ using the two approaches and
isochrones from
four different stellar evolution codes: Geneva (Schaller et al.\ 1992),
Bertelli (Bertelli et al.\ 1994), Claret (Claret 1995), and Yale
(Demarque et al.\ 1996). The results of this comparison are given in the
first four lines of Table 2.

The last two lines of Table 2 demonstrate the effects of changing the
stellar helium abundance and metallicity. The next-to-the-last line
indicates that helium-rich models give slightly lower values of the mass
and radius.  The effect of changing the metallicity, with the helium
scaled by the enrichment law, is illustrated by the last line. We
therefore conclude that if Z = 0.019 is adopted for the solar
metallicity (Anders \& Grevesse 1989), all the results for the Z = 0.02
models should be reduced by about 0.01 in both mass and radius.

Note that all four sets of evolutionary models and the two approaches
yielded similar results, with only small differences.  The mass
estimates are systematically smaller for the results based on the
observational \Mv\ vs $B-V$ plane, by about $0.05 \,\msun$.  The radius
estimates are also smaller, by about $0.03 \,\rsun$. Note, however, that the good
agreement between the different models may be misleading, and that systematic
errors are likely to be larger.  Based on all these considerations, we
adopt for our best estimate of the mass and radius of HD~209458
$1.1\pm0.1 \,\msun$ and $1.3\pm0.1\,\rsun$. The uncertainty estimates are
somewhat arbitrary, and are based mainly on the assumed uncertainty in
\Teff, \logg, \Mv\ and $B-V$.

Using ELODIE and CORAVEL cross-correlation dip widths, we infer
the projected rotational velocity (calibration by Queloz et al. (1998)
for ELODIE and by Benz \& Mayor (1984) for CORAVEL). The results are
$v\sin i = 4.4 \pm 1$ \kms\ for CORAVEL and $v\sin i = 4.1 \pm 0.6 $ \kms\
for ELODIE. 

\section{PLANETARY PARAMETERS}

C00 present preliminary estimates of the planetary radius, $R_{p}$, and
orbital inclination, $i$, based on initial estimates of \Ms, \Rs, and the
$R$-band limb-darkening parameter $c_{R}$.  We now present values for
these quantities based on the more accurate analysis of the stellar
parameters presented in this Letter.  In addition, we present estimates
of the uncertainties that combine the effects both due to the
uncertainties in the stellar parameters, and due to the level of
precision in the photometric measurements of the transit.  All
uncertainties presented below correspond to 1-$\sigma$ confidence
levels.

Using the calculated limb darkening coefficients presented
in Claret, D\'{i}az-Cordov\'{e}s, \& Gim\'{e}nez (1995), we adopted a
value of $c_{R} = 0.56 \pm 0.05$, based on the values for \Teff\ and
\logg\ derived in the previous section. As described in C00, we then
calculated the $\Delta {\chi}^2$ of the photometric points for the model
light curve as a function of $R_{p}$ and $i$, using the revised values
of \{\Ms,\,\Rs,\,$c_{R}$\} presented here.  To evaluate the uncertainty
in the derived values of $R_{p}$ and $i$, we calculated the $\Delta
{\chi}^2$ for all combinations of the stellar parameters at 1-$\sigma$
above and below their respective best fit values.  We then assign
1-$\sigma$ error bars based on the intervals which are excluded with
this confidence for all these combinations.

We find $R_{p} = 1.54 \pm 0.18 \,R_{\rm Jup}$ and $i = 85\fdg2 \pm
1\fdg4$.  The primary mass and the inclination imply (see Table 1) 
the planetary mass to
be

\begin{equation}
M_p=0.69\pm 0.05 \,\MJ \ .
\end{equation}

>From the planetary radius and mass we calculate the density, surface
gravity, and escape velocity to be 
$\rho = 0.23 \pm 0.08 \, {\rm g \,cm^{-3}}$, 
$g = 720 \pm 180 \, {\rm cm \, s^{-2}}$, 
and $v_{e} = 40 \pm 4\, {\rm km \, s^{-1}}$. 

In the interpretation of the transit curve of HD~209458, the dominant
uncertainty in the planetary parameters is due to the uncertainty in the
stellar radius, rather than the observational uncertainty in the
photometric points.

The planetary radius found here is larger and the orbital inclination is
smaller than the values presented in C00.  This is due primarily to the
fact that the value of the stellar radius found here ($1.3\,\Ro$) is
larger than the one assumed in the initial analysis ($1.1\,\Ro$). A larger
star requires a larger planet crossing at a lower inclination to fit the
same photometric data.
The results presented here and in C00 are based on the analysis of the
detailed observed transit lightcurve. Henry et al. (2000), who did not
observe the full transit, assumed an inclination of 90\degr\ for the
orbital inclination, and a stellar radius and mass of $1.15\,\Ro$ and
$1.03\,\Mo$. With these
assumptions they derived a planetary mass and radius of $R_p=1.42\pm
0.08 \,\RJ$ and $M_p=0.62\,\MJ$.

\section{DISCUSSION}

The spectroscopic orbit, when combined with the inclination derived from
transits, enables us to derive the planetary mass directly. This demonstrates
the power of combining spectroscopy and photometry for transiting planets
(C00; Henry et al.\ 2000). To derive masses for
non-transiting planets that have spectroscopic orbits, we are forced to
turn to other approaches for determining the orbital inclination, such as
astrometry (e.g., Mazeh et al. 1999; Zucker \& Mazeh 2000).

\hd\ is the first extrasolar planet whose orbital inclination is known
with high precision. In principle we might be able to derive the
inclination of the stellar rotational axis, if we could obtain the
stellar rotational period from photometric observations. Together with
the projected rotational velocity and the stellar radius derived here,
this will enable us for the first time to check the assumption that the
stellar rotation is aligned with the orbital motion for such
short-period systems. With $v\sin i = 4.2 \pm 0.5$ \kms, alignment
implies a rotational period of $P = 15.7 \pm 2.4$ days.

One unique feature of the transit technique is its ability to derive the
planetary radius.  As described in the review by Guillot (1999) and the
references listed therein, the radius of an extrasolar giant planet is
determined by its mass, age, degree of insolation, and composition.  Now
that an accurate measurement of the planetary radius has been made, it
should be possible to infer specifics of the planetary composition,
since the mass and insolation are known, and the age can be reasonably
constrained based on the value of the age of the star determined above.
More specifically, as described in Guillot (1999), it may be possible to
calculate the amount of heavy elements for a given hydrogen/helium
ratio, and to infer the presence or
absence of certain atmospheric grains.

These very fascinating possibilities, together with the analysis of other
known short-period extrasolar planets, like 51 Peg, $\tau$ Boo, and
$\upsilon$ And (e.g., Ford, Rasio \& Sills 1999), promise us new insights
to the formation and evolution of these systems.

\acknowledgments 

We give special thanks to Robert L. Kurucz for generously sharing his
expertise and procedures for analyzing stellar spectra.  We are grateful
to Karin Sandstrom for her help with the abundance analysis and to the
referee for his useful comments.  This work was supported by the
US-Israel Binational Science Foundation through grant 97-00460, the
Israeli Science Foundation, the Swiss NSF (FNRS) and the French NSF
(CNRS).

 \section*{REFERENCES}
 \begin{description}

 \item Anders, E., \& Grevesse, N. 1989, \gca, 53, 197

 \item Baranne, A., Queloz, D., Mayor, M., Adrianzyk, G., Knispel, G.,
 Kohler, D., Lacroix, D., Meunier, J.-P., Rimbaud, G., \& Vin, A. 1996,
 A\&AS, 119, 373

 \item Benz W., \& Mayor M., 1984, A\&A, 138, 183

 \item Bertelli, G., Bressan, A., Chiosi, C., Fagotto, F., \& Nasi, E. 1994,
 A\&AS, 106, 275

\item Brown, T.~M., \& Kolinski, D. 1999,\\
 http://www.hao.ucar.edu/public/research/stare/stare.html
 
\item Carney, B.~W., Laird, J.~B., Latham, D.~W., \& Kurucz, R.~L. 1987,
 \aj, 94, 1066

 \item Charbonneau, D., Brown, T.~M., Latham, D.~W., \& Mayor, M. 2000,
 ApJL, 529, L45

 \item Claret, A. 1995, A\&AS, 109, 441

 \item Claret, A., D\'{i}az-Cordov\'{e}s, J., \& Gim\'{e}nez, A. 1995, A\&AS, 
 114, 247

 \item Demarque, P., Chaboyer, B., Guenther, D., Pinsonneault, M.,
 Pinsonneault, L., \& Yi, S. 1996, Yale Isochrones 1996 in "Sukyoung Yi's WWW
 Homepage".

 \item Edvardsson, B., Andersen, J., Gustafsson, B., Lambert, D. L., Nissen,
P.E., \& Tomkin, J., 1993, A\&A 275, 101

 \item ESA 1997, The Hipparcos and Tycho Catalogues, ESA SP-1200

 \item Ford, E.~B., Rasio, F.~A., \& Sills, A. 1999, \apj, 514, 411

 \item Guillot, T. 1999, Science, 286, 72

 \item Henry, G., Marcy, G.~W., Butler, R.~P., \& Vogt, S.~S. 2000,
   \apjl, 529, L41

 \item Kurucz, R.~L., Furenlid, I., \& Brault, J. 1984, Solar Flux Atlas from
 296 to 1300 nm, National Solar Observatory Atlas No. 1

 \item Latham, D.~W. 1985, in Stellar Radial Velocities, IAU Coll. 88,
 ed. A.~G.~D. Philip \& D.~W. Latham, (Schenectady: L. Davis Press),
 p. 21

 \item Latham, D.~W. 1992, in Complementary Approaches to Double and Multiple
 Star Research, IAU Coll. 135, ed. H.~A. McAlister
 \& W.~I. Hartkopf, ASP Conference Series, Vol. 32, p. 110.

\item Latham, D.\ W.\, Charbonneau, D., Brown, T.\ M., Mayor, M., \& Mazeh,
T. 1999, IAUC, 7315

 \item Latham, D.~W. 2000, in Bioastronomy 99: A New Era in the Search for Life
in the Universe, ed. G. Lemarchand and K. Meetch, ASP Conference
Series, in press

 \item Marcy, G.~W., \& Butler, R.~P. 1992, \pasp, 104, 270

 \item Mayor, M. 1980, \aap, 87, L1

 \item Mayor, M. 1985, in Stellar Radial Velocities, IAU Coll. 88,
 ed. A.~G.~D. Philip \& D.~W.~Latham, (Schenectady: L. Davis Press),
 p. 35

 \item Mayor, M., \& Queloz, D. 1995a, in Cool Stars, Stellar Systems 
and the Sun, 9th Cambridge workshop, ed. R. Pallavicini \& A.~K. Dupree, ASP
Conference Series, Vol. 109, p.35

 \item Mayor, M., \& Queloz, D., 1995b, Nature, 378, 355

 \item Mazeh, T., Zucker, S., Dalla Torre, A., \& van Leeuwen, F., 1999,
  \apjl, 522, L149

\item Pont, F., 1997, Ph.D. Thesis, Geneva Observatory

\item Queloz, D., Allain, S., Mermillod, J.-C.,
       Bouvier, J., \& Mayor, M. 1998, A\&A, 335, 183

\item Queloz, D., Mayor, M., Weber, L.,
  Bl\'echa, A., Burnet, M., Confino, B., Naef, D., Pepe, F., Santos, N., \&
  Udry, S. 1999, A\&A, in press

\item Robichon, N., \& Arenou, F.\ 2000, A\&A, in press 

\item Schaller, G., Schaerer, D., Meynet, G., \& Maeder, A. 1992, A\&AS,
96, 269

\item Udry S., Mayor M., Queloz, D., Naef D., 
  Santos N.C., 1999a, VLT Opening Symposium, ``From extrasolar
  planets to brown dwarfs'', ESO Astrophys. symp. Ser., in press

\item Udry, S., Mayor, M., Naef, D., Pepe, F., Santos, N.~C.,
Queloz, D., Burnet, M., Confino, B., \& Melo, C., 1999b, A\&A, submitted

\item Vogt, S.~S., et al. 1994, Proc. Soc. Photo-Opt. Instrum. Eng.,
  2198, 362

\item Zucker, S., \& Mazeh, T. 1994, \apj, 420, 806
\item Zucker, S., \& Mazeh, T. 2000, ApJL, in press

\end{description}

\begin{center}
FIGURE CAPTION
\end{center}

\figcaption{Radial velocities of HD209458,
 plotted as a function of orbital phase for the solution detailed
  in Table~1. The measurements of the three observational programs are 
represented by different symbols.}

\begin{table}
\caption{Orbital Solution for HD~209458.}
\vskip 0.2truecm
\begin{tabular}{lcl}
\hline
Period     &   $3.52433 \pm 0.00027$      &   days   \\
$\gamma$   &   $-14.7652 \pm 0.0016$      &   \kms\   \\
K          &   $ 85.9 \pm 2.0      $      &   \ms\    \\
e          &      0                       &   FIXED  \\ 
$T_c$  &   $2,451,430.8238 \pm 0.0029$    &   HJD    \\
$M_p\sin i$ &
$0.685 \pm 0.018\ ({M_{\ast}/{1.1M_{\odot}}})^{2/3}$& \MJ\\
\hline
$\Delta_{H-C}$   &        $+0.2\pm 4.7$        &   \ms\    \\ 
$\Delta_{E-C}$   &        $+0.5\pm 5.1$        &   \ms\    \\
$\sigma_H$  &        $13.8$              &   \ms\    \\
$\sigma_E$  &        $25.1$              &   \ms\    \\
$\sigma_C$  &        $17.6$              &   \ms\    \\
\hline
\end{tabular}
\end{table}

\begin{table} 
\caption{The Mass and Radius of \hd} 
\vskip 0.2truecm
\begin{tabular}{lllcccccc} 
\hline
\multicolumn{3}{c}{Model} & \multicolumn{3}{c}{\logg\ vs. \Teff} &
        \multicolumn{3}{c}{\Mv\ vs. $B-V$} \\
\hline
Code    &\ \ Z  &\ \ Y &  Age &  \Ms &   \Rs &    Age &  \Ms  & \Rs  \\
        &      &     & (Gyr)&  (\Mo) & (\Ro) &(Gyr) & (\Mo)&  (\Ro)\\
\hline
Geneva  &  0.02&  0.30 & 4.6  & 1.15 & 1.33  &  6.3 & 1.08  & 1.29 \\
Bertelli&  0.02&  0.27 & 5.0  & 1.11 & 1.31  &  4.0 & 1.09  & 1.30  \\
Claret  &  0.02&  0.28 & 5.3  & 1.12 & 1.31  &  7.9 & 1.05  & 1.27  \\
Yale    &  0.02&  0.27 & 5.7  & 1.11 & 1.31  &  7.3 & 1.06  & 1.28  \\
\hline
Yale    &  0.02&  0.30 & 6.0  & 1.05 & 1.27  &  7.7 & 1.01  & 1.25  \\
Geneva  & 0.008&  0.264& 9.8  & 0.94 & 1.20  &12.3\ & 0.91  & 1.30  \\
\hline
\end{tabular}
\end{table}

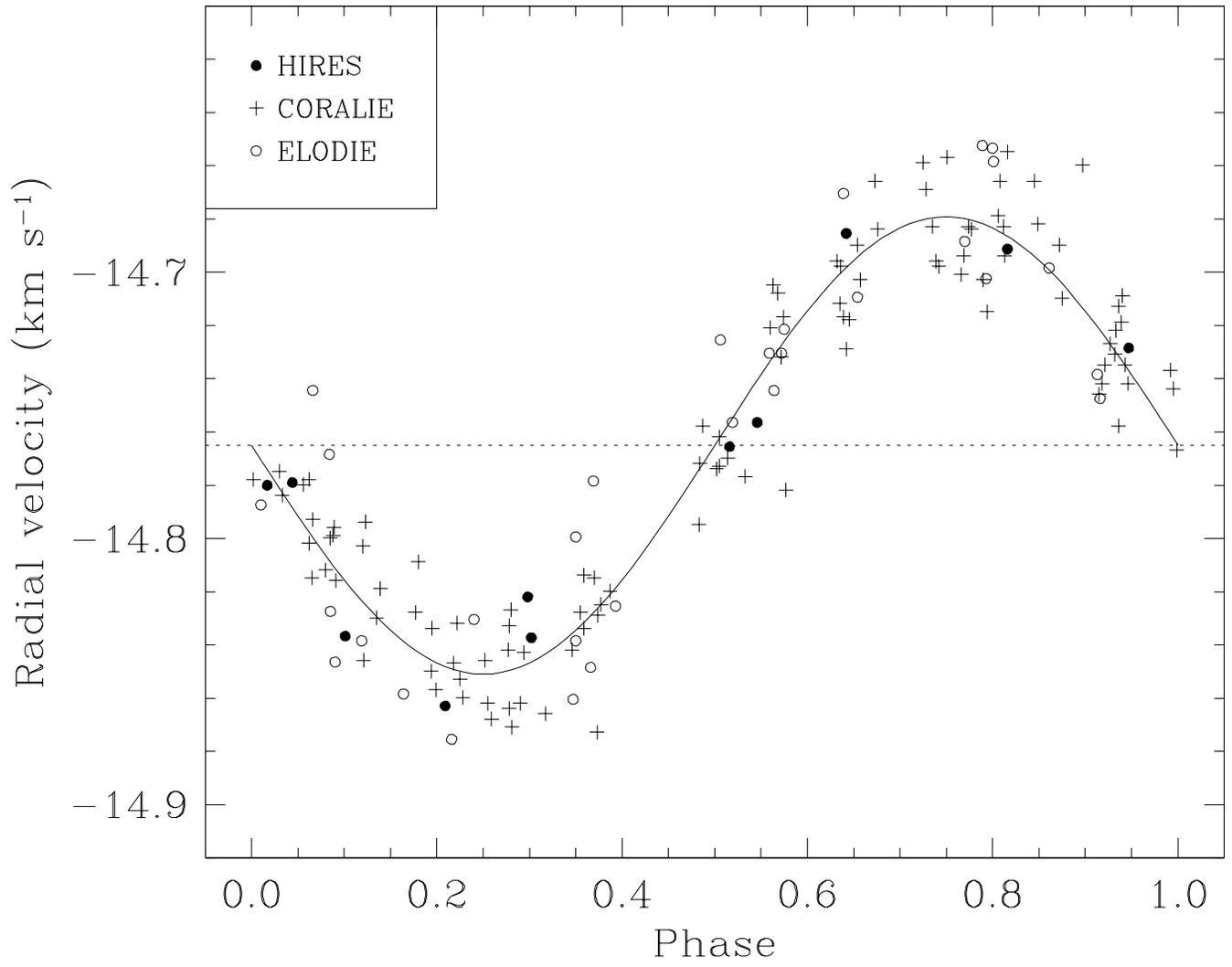
\end{document}